# Electrically Detected Magnetic Resonance Modeling and Fitting: An Equivalent Circuit Approach

D. M. G. Leite[1,*], A. Batagin-Neto[2], O. Nunes-Neto[2], J. A. Gómez[3], C. F. O. Graeff[1,2]

[1]DF-FC, UNESP - Univ Estadual Paulista, Av. Eng. Luiz Edmundo Carrijo Coube, 14-01, 17033-360, Bauru, SP, Brazil
[2]UNESP - Univ Estadual Paulista, POSMAT - Programa de Pós-Graduação em Ciência e Tecnologia de Materiais, Av. Eng. Luiz Edmundo Carrijo Coube, 14-01, 17033-360, Bauru, SP, Brazil
[3]Departamento de Física, FFCLRP-USP, Av. Bandeirantes 3900, 14040-901, Ribeirão Preto, SP, Brazil
*dmgleite@fc.unesp.br

Continuous-wave electrically detected magnetic resonance (cw-EDMR or just EDMR) and its variants are powerful tools to investigate spin-dependent processes in materials and devices. The use of quadrature detection improves the quality and selectivity of EDMR analysis by allowing, for example, the separation of individual resonant spin lines. Here we propose an equivalent circuit model in order to better understand the EDMR quadrature signal in a variety of different situations. The model considers not just the electrical response of the sample but of cables and measuring circuit and its influence on the resulting spectral lines. Recent EDMR spectra from $Alq_3$ based OLEDs, as well as from a-Si:H reported in the literature, were successfully described by the model. Moreover, the model allows the implementation of a fitting routine that can be easily used to determine accurate values of crucial parameters such as $g$-factor and linewidth of the resonant lines.

## I. INTRODUCTION

Electronic magnetic resonance techniques (EMRT) are powerful tools widely employed in material research and device characterization. In general these techniques are based on detecting changes in a given observable when electron spin resonance condition is reached, and allows the investigation of intrinsic properties of charge carrier's spin, as well as its interaction with the local environment[1-4]. Electron spin resonance (ESR), electrically detected magnetic resonance (EDMR) and optically detected magnetic resonance (ODMR) are the most representatives EMRT[5-11].

In what concerns electronic device characterization, EDMR is the most appropriate technique, since experiments can be done in real devices under working conditions[12-14]. During charge transport or recombination, paramagnetic or ESR active spin pairs are commonly formed in the device. Since transitions involving these centers, called precursor pairs, are often governed by spin selection rules, measurable changes in conductivity can be induced by magnetic resonance, resulting in an EDMR signal. In this sense it is a very sensitive and selective tool to investigate the transport or recombination mechanisms responsible for the device operation.

Indeed, different spin-dependent processes (SDPs) have been measured by EDMR. They have been commonly associated to spin-dependent scattering, spin-dependent tunneling and spin-dependent trapping/recombination mechanisms[15]. In the particular case of disordered and organic materials these mechanisms have been discussed in terms of bipolarons and excitons formation[13,16,17], polaronic and excitonic recombination[15,18] or triplet-triplet annihilation[19].

Since these processes involve precursor pairs formed from electrons (or holes), electron/hole pairs, or excitons, at different nanoscopic environments, it is expected that they present dissimilar spectroscopic characteristics. In this sense EDMR spectra can be considered as a complex combination of distinct resonant spin lines (RSL).

Dersch *et al.*[20] were one of the first to explore the use of phase analysis of the EDMR signal, in order to discriminate SDPs in a-Si:H. By tuning the detection phase, they identified and isolated two SDPs attributed to tail states electrons recombining with dangling bonds and hopping of holes in tail states. After that, many other research groups have explored phase analysis in order to extract more and potentially better physical information from EDMR experiments[21-25].

Recently, Lee *et al.*[26] presented a thorough study on ODMR and EDMR experiments. It was demonstrated that the phase response of the device is strongly dependent on intrinsic transition rates of the RSL, as well as on the experimental parameters used during data acquisition. In their work the relative intensity of in-phase and out-of-phase components of the EDMR signal were evaluated using rate equations. Nevertheless their study did not concern lineshape analysis and its use on the discrimination of different RSL as will be discussed here.

In order to fully address the EDMR experiment, the measuring electric circuit must be considered. Wimbauer *et al.*[27] proposed a RC equivalent circuit composed of two photo-resistors responding with different phases to a modulated light excitation to simulate EDMR in multilayer devices. Contrary to what is proposed here, in his approach the focus was on simulating the amplitude and phase of the various components of the EDMR signal and not lineshape or other spectroscopic characteristics such as $g$-factor.

In the present study we extend Wimbauer's proposition. For this purpose, the RSL are modeled using Pseudo-Voigt lineshapes, with well-defined g-factors. In our proposal magnetic field modulation effects are an integral part of the description. The model allows the simulation of different phase dependent EDMR signals providing valuable information regarding its origin and dependence with experimental conditions.

The work is organized as follow. In section II, the equivalent circuit model employed and phase detection details are presented, followed by section III where a brief discussion of some implications of the circuit model, as well as examples regarding out-of-phase signals are presented. In section IV the methodology is applied to experimental results found in the literature. Finally in section V the main conclusions are presented. In addition, line-broadening effects induced by excessive magnetic field modulation are presented and discussed in appendix A, and a fitting software is presented in Appendix B.

## II. THE CIRCUIT MODEL

EDMR (or cw-EDMR) spectra are ordinarily obtained by monitoring conductivity changes by phase sensitive detection (PSD). For that purpose, the magnetic field is modulated, creating a small AC in the measuring circuit at the resonance condition, which is much smaller than the sample DC current, typically AC≤$10^{-4}$ DC.

Given the sample's characteristics and the measuring circuit used, the EDMR setup configures an electrical circuit with a variety of elements that in most cases is described as a RC circuit. In special, the capacitance makes the EDMR circuit a low-pass filter. When double channel Lock-in amplifier is used the output signal is divided in two channels, $X$ and $Y$ in phase quadrature. It is common practice that the signal in the $X$ channel is maximized and used as the 0° detection phase. Consequently, the signal in the $Y$ channel (in 90°) is zero or very small.

### A. The equivalent circuit

Equivalent circuits are an efficient way of obtaining relevant physical information from complex systems. For example, it can be used to simulate the impedance response of double or multi junction devices such as light emitting diodes (LEDs), organic LEDs (OLEDs) and solar cells[28-31].

In our model, the sample or device is described as a parallel RC circuit with a constant capacitance $C_S$ and a rheostat $R$ that depends on the applied magnetic field $H$. The dependence of $R(H)$ basically emulates the conductivity changes under resonance as will become clearer in the following discussion. Figure 1(a) shows the schematic representations of the proposed equivalent circuit. $V_0$ is the external applied voltage; $R_M$ is the measuring resistance used to convert the current into measured voltage $V_M$. $C_C$ is the circuit capacitance, mainly due to cables and connections.

Figure 1(b) shows the simplified version of the proposed equivalent circuit by converting the sample resistance $R$ to conductance $G=1/R$, and associating the sample capacitance ($C_S$) with the measuring circuit capacitance ($C_C$) as the equivalent capacitance $C=C_S+C_C$, which is converted to the circuit susceptance $B=\omega C$. Here $\omega$ stands for the angular frequency of the modulation induced AC. Notice that the AC is observed only close to resonance.

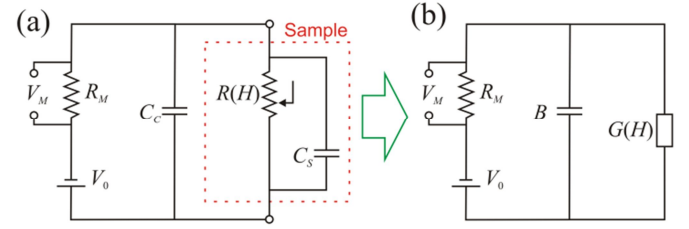

**Figure 1.** (a) Squematic representation of the real measuring circuit with sample as equivalent parallel RC circuit and (b) the correspondent simplified version used to develop the model.

In the proposed model, $G$ has a constant term $G_0$ plus a magnetic field dependent term represented by a pseudo-Voigt profile function:

$$G = G_0 + G_H\left[\alpha\left(2^{-4\Omega^2}\right) + (1-\alpha)\left(\frac{1}{1+4\Omega^2}\right)\right], \text{with} \quad (1a)$$

$$\Omega = \frac{H-H_{RES}}{\Delta H_{1/2}}, \text{and} \quad (1b)$$

$$H_{RES} = \frac{h\nu}{\mu_B g}, \quad (1c)$$

where $\Delta H_{1/2}$ the full width at half maximum of the resonant line, $h\nu$ the microwave energy, $\mu_B$ the Bohr magneton, $g$ the g-factor, and $G_H$ the maximum conductance shift (positive or negative) in the resonant condition.

In an EDMR experiment usually $h\nu$ is constant so the magnetic field is swept linearly around a central field $H_0$ close to the resonant condition. As discussed previously a modulation field with angular frequency $\omega=2\pi f$, and amplitude $H_m$ is used for PSD resulting in:

$$H(t) = \underbrace{H_0 + \Delta H\left(\frac{t}{\Delta t} - \frac{1}{2}\right)}_{H_L} + \underbrace{H_m \sin(\omega t)}_{H_{MOD}}, \quad (2a)$$

$$H = H_L + H_{MOD}, \quad (2b)$$

where $\Delta H$ and $\Delta t$ represent the amplitude of the field sweep and the sweep time respectively. For simplicity we define $H_L$ as the field component, which is swept, and $H_{MOD}$ as the modulated component. The time dependence of $H_L$ and $H_{MOD}$ was omitted in the second equation.

### B. Signal processing and detection

It is easy to realize that $G$ has no simple time dependence, and thus it is convenient to use Fourier expansion:

$$G(t) \approx c_0 + c_1 \sin(\omega t + \phi) + \Lambda, \quad (3)$$

where $c_i$ are the Fourier coefficients and $\Lambda$ stands for the higher frequency terms. Here we define $\phi$ as a constant phase shift from the reference signal.

As in the exhaustively studied ESR case, the Fourier coefficients can be determined via Taylor expansion[32]. For low values of $H_m$ in comparison to $\Delta H_{1/2}$ ($H_m/\Delta H_{1/2}$<<1), which is the most common situation in EDMR and ESR experiments, it is possible to use the following approximations for the dependence of $c_0$ and $c_1$ with $H_L$:

$$c_0(H_L) \approx G(H_L), \qquad (4)$$

$$c_1(H_L) \approx \frac{H_m}{2} \frac{dG}{dH}\bigg|_{H=H_L} = \frac{H_m}{2} \frac{dG}{dH}(H_L). \qquad (5)$$

Remember that $c_1$ is measured in PSD. Notice that when $H_m \sim \Delta H_{1/2}$, the approximation of $c_1$ given by Eq. 5 is no longer valid, see Appendix A.

For $R_M \ll R$, which is normally the case, the DC component of $V_M$ can be approximated to:

$$V_{DC}(H_L) = V_0 R_M G(H_L). \qquad (6)$$

While the output signal of the PSD can be written as:

$$V_{AC}(H_L) = S(H_L) \angle \Phi(H_L), \qquad (7)$$

where:

$$\Phi(H_L) = \phi + \tan^{-1}(B/G(H_L)) \qquad (8)$$

represents the phase of the measured signal, and

$$S(H_L) = \beta V_0 R_M c_1(H_L) \qquad (9)$$

is its magnitude. $S$ can assume either positive or negative value at a given $H_L$ due to the definition of $c_1$, with $\beta$ representing the attenuation factor of the equivalent RC circuit, given, in a first approximation, by:

$$\beta = [1 + (B/G_0)^2]^{-1}. \qquad (10)$$

The quadrature signal decomposed in $X$ and $Y$ channels for the fundamental frequency is then given by:

$$X(H_L) = S(H_L) \cos(\Phi(H_L) - \theta), \qquad (11)$$

$$Y(H_L) = S(H_L) \sin(\Phi(H_L) - \theta), \qquad (12)$$

where $\theta$ is an offset angle.

For the following discussions and applications, we always set the detection angle $\theta$ equal or very close to the average value of $\Phi(H_L)$ over the whole $H_L$ sweep. In this way, the EDMR signal is mainly in the $X$ channel, with a small component at the $Y$ channel.

## III. DISCUSSION

The microscopic origin of the EDMR signal with non-vanishing component at $Y$ channel has been discussed in terms of different mechanisms involving, at least, two SDPs or one SDP with a spin pair with different phases. Indeed, SDPs or spin pairs generally experience distinct chemical environment with different spin-orbit or hyperfine couplings as well as spin-lattice relaxation times. These characteristics can lead to EDMR spectra composed by RSL with different resonant fields ($H_{RES}$), different linewidths ($\Delta H_{1/2}$) and different response phases ($\phi$).

For example, Dersch et al.[20] distinguished two independent SDPs in their EDMR study of a-Si:H. A narrow line with $g \sim 2.005$ attributed to spin-dependent recombination of localized band-tail electrons with holes in dangling bonds and a broad line at $g \sim 2.01$ attributed to spin-dependent hopping of localized band-tail holes. These two RSL were found to be dephased by 15° when using a modulation frequency of $f=1$ kHz. Notice that for the hopping related signal the two spins are basically in the same nanoscopic environment in the mobility edge of a-Si:H valence band, while the recombination related signal is coming from distinct spins. The electron is occupying the mobility edge of a-Si:H conduction band, while the hole is in a dangling bond. Graeff et al.[22,25,33] separated the contribution of each spin that formed the exciton precursor pair of Alq$_3$ based OLEDs. In this case only one SDP is responsible for the EDMR signal, and again two distinct spins are participating, an electron in the LUMO of Alq$_3$ (or a dopant molecule), and a hole in the HOMO of Alq$_3$ or α-NPD. Contrary to the case of electron recombination in a-Si:H, the two spins give rise to two RSL. The reason why in one case only one RSL is observed and in the second two is not in the scope of this article. However it is important to notice that this effect may be induced by the microwave energy, or magnetic field used in resonance. As the magnetic field is increased there is a higher separation in resonant field between spins with different g-factors. In this context, Dersch's work was made in X-band, 9 GHz, while Graeff's in K-band, 24 GHz.

In the following, the influence of $H_{RES}$, $\Delta H_{1/2}$ and $\phi$ on the simulated quadrature EDMR signal is presented in two different regimes depending on the modulation frequency used.

### A. Low modulation frequency

In the low modulation frequency (LMF) regime $B/G_0 < 0.01$, which is typically used in EDMR experiments, the signal phase of a single RSL assumes a constant value, $\Phi(H_L) \rightarrow \phi$ in Eq. 8. If only indistinguishable spins are present, for example in spin dependent hopping, the EDMR signal is solely projected in the $X$ channel, and thus no signal is observed in the $Y$ channel, see Eq. 11 and Eq. 12 when $\theta = \phi$.

However, as discussed previously, the EDMR signal is commonly composed by two RSL. Thus, if one considers two independent resonant spin lines, RSL$^a$ and RSL$^b$, responding with different phases, $\phi^a$ and $\phi^b$ with respect to the reference, a signal will be observed in the $Y$ channel under certain conditions as will be discussed in the following.

In the present circuit model, the presence of two RSLs is represented by splitting the field dependent term in Eq. 1 into two summing terms with independent parameters: $G_H^a$, $\alpha^a$, $H_{RES}^a$ and $\Delta H_{1/2}^a$ for the first RSL, and $G_H^b$, $\alpha^b$, $H_{RES}^b$ and $\Delta H_{1/2}^b$ for the second. This approach does not rely necessarily in any specific association of the sample resistance/conductance: in series or in parallel.

Figure 2 shows the simulated quadrature EDMR signals for two representative conditions, in (a) $H_{RES}^a > H_{RES}^b$ with $\Delta H_{1/2}^a = \Delta H_{1/2}^b$, and in (b) $\Delta H_{1/2}^a > \Delta H_{1/2}^b$ with $H_{RES}^a = H_{RES}^b$. In all cases $\Delta \phi = 1°$, $\alpha^a = \alpha^b$ and $G_H^a = G_H^b > 0$. For each condition seven curves are plotted for $\theta$ varying from $\phi^a$ to $\phi^b$. Notice that $\theta \approx (\phi^a + \phi^b)/2$ minimizes the signal amplitude in the $Y$ channel, and that the small changes in $\theta$ from $\phi^a$ to $\phi^b$ do not affect the signal on the $X$ channel.

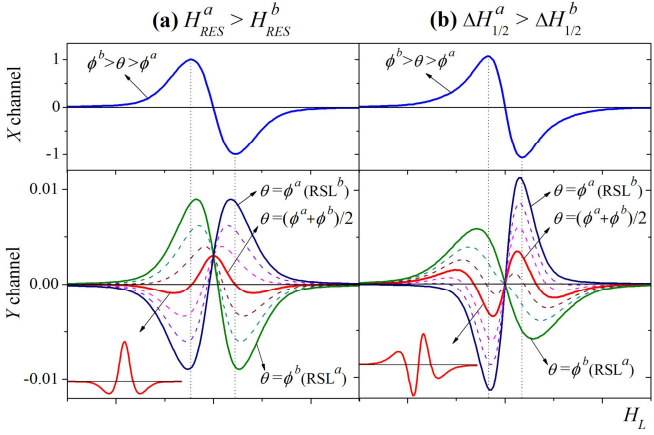

**Figure 2.** Simulated EDMR signal in the *X* and *Y* channels of a dual phase lock-in amplifier for $G_H^a = G_H^b > 0$, $B = 0$, $\alpha^a = \alpha^b$, $\Delta\phi = 1°$, and: (a) $H_{RES}^a > H_{RES}^b$ and $\Delta H_{1/2}^a = \Delta H_{1/2}^b$; (b) $H_{RES}^a = H_{RES}^b$ and $\Delta H_{1/2}^a > \Delta H_{1/2}^b$. For each condition seven curves are plotted for $\theta$ varying from $\phi^a$ (showing the individual signal of RSL$^b$ in the *Y* channel) to $\phi^b$ (showing the individual signal of RSL$^a$ in the *Y* channel). Notice that small changes in $\theta$, do not affect the signal in the *X* channel.

It is important to emphasize that both cases shown in Figure 2 are only representative, i.e. it is unlike to have RSLs with exactly the same *g*-factors or same line-widths. However the quadrature signals of Figure 2 with $\theta \approx (\phi^a + \phi^b)/2$ are useful to discriminate whether, in real EDMR spectra, the difference on $H_{RES}$ or on lineshape (represented here by $\Delta H_{1/2}$) is the dominant effect.

In both cases described in Figure 2, the individual RSLs are isolated in the *Y* channel by shifting $\theta$ from $\phi^a$ to $\phi^b$: RSL$^b$ (blue solid line) and RSL$^a$ (green solid line). This procedure was used for example in Refs. [20] and [22] for this purpose. Note also that there is no visible change in the signal in the *X* channel in all conditions displayed due to the low value of $\Delta\phi = 1°$. For higher values of $\Delta\phi$ not shown here, significant changes in the *X* channel are observed.

The special case of Figure 2(a) with $\theta = (\phi^a + \phi^b)/2$ is reproduced in Figure 3 in order to elucidate the dependence of the signals in *X* and *Y* channels on $\Delta\phi$ (Figure 3(a)) and $\Delta H_{RES} = H_{RES}^a - H_{RES}^b$ (Figure 3(b)) given in terms of the $\Delta H_{1/2}$ used for both lines.

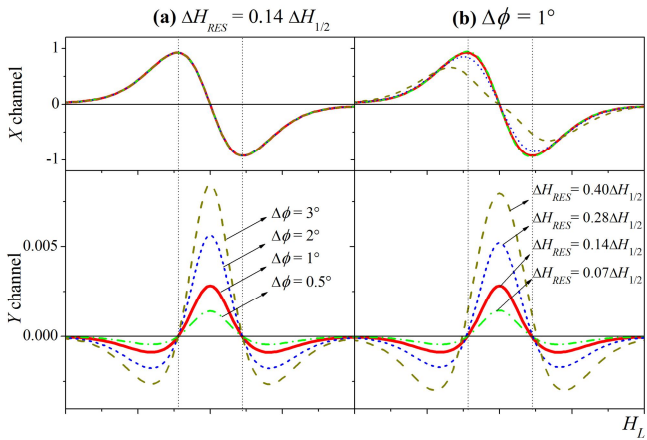

**Figure 3.** Simulated EDMR signal in the *X* and *Y* channels of a dual phase lock-in amplifier for a similar condition as in Figure 2(a). In (a) $\Delta\phi$ is varied from 0.5° to 3° and in (b) $\Delta H_{RES}$ is varied from $0.07\Delta H_{1/2}$ to $0.4\Delta H_{1/2}$.

Notice that the increase on either $\Delta\phi$ (Figure 3(a)-bottom) or $\Delta H_{RES}$ (Figure 3(b)-bottom) increases the signal amplitude in the *Y* channel. However, increasing $\Delta\phi$ does not affect the lineshapes (Figure 3(a)). On the other hand, both lineshapes in the *X* and *Y* channels are affected by increasing $\Delta H_{RES}$ (Figure 3(b)) specially when $\Delta H_{RES}$ approaches $0.5\Delta H_{1/2}$.

The model allows the analysis of other more complicated situations not shown here.

## B. High modulation frequency and one resonant spin line

In the special case of high modulation frequencies (HMF), where $B/G_0 > 0.01$, the current through the capacitor *C* in Figure 1(b) is responsible for two main features: (*i*) it attenuates the measured signal, *S*, acting as a low-pass filter, represented by $\beta$ in Eq. 9; and (*ii*) it promotes dynamic phase-shifts on the signal, see Eq. 8, and thus a signal in the *Y* channel in any condition, contrary to what was just described in the LMF regime.

The most important feature for the quadrature detection is that $\theta$ represents a constant projection angle in Eqs. 11 and 12, while $\Phi(H_L)$ in Eq. 8 is a field dependent parameter, which varies from $\Phi_0$ to $\Phi_P$ in the same way that $G(H_L)$ varies from $G_0$ to $G_0 + G_H$ in the resonance condition. In the case of only one RSL with constant phase $\phi$, it implies that:

$$\Phi_0 = \phi + \tan^{-1}(B/G_0), \qquad (13)$$

$$\Phi_P = \phi + \tan^{-1}(B/(G_0 + G_H)). \qquad (14)$$

Thus, even if only a single RSL is present, there is no $\theta$ in which the EDMR signal has no projection at the *Y* channel. Figure 4(a) shows an example of a quadrature signal for an enhancing single RSL EDMR signal ($G_H > 0$) where $B/G_0 = 0.01$, $G_H/G_0 = 0.02$. The amplitudes are normalized by the maximum in the *X* channel. The signal in the *Y* channel is shown using different $\theta$ from $\Phi_0$ to $\Phi_P$. For this or any similar condition, variation in $\theta$ in this range does not affect the signal in the *X* channel.

Notice that in Figure 4(a) the minimum amplitude in *Y* is reached when $\theta \approx (\Phi_0 + \Phi_P)/2 = \Phi_M$ as expected. In this case a non-ordinary line-shape that resembles the third derivative of *G* or the second derivative of the signal in the *X* channel is observed.

Figure 4(b) shows the simulated EDMR signal amplitude at channels *X* and *Y* as a function of $B/G_0$, using $\theta = \Phi_M$ and $G_H/G_0 = 0.02$. The dependence of the signal in the *X* channel is of a low pass filter as expected, Eq. 10. The behavior of the signal in the *Y* channel is dominated by $\Phi_0 - \Phi_P$ but $\beta$ plays a relevant role. The ratio of the amplitude of the signals in the *Y* and *X* channels reaches a maximum when $B \approx G_0$ (or equivalently $\Phi_0 \approx \phi + 45°$), close to the attenuation threshold, i.e., when the signal in the *X* channel starts to be attenuated. To give a numerical example, this situation is reached at $f = 10$kHz for $R = 1/G_0 = 100$k$\Omega$ and $C = 1$nF.

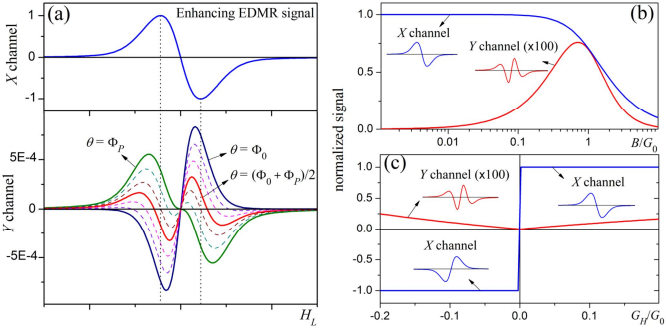

**Figure 4.** (a) Simulated EDMR signal for $G_H>0$, $B/G_0=0.01$ and $G_H/G_0=0.02$. The amplitude is normalized by the maximum of the signal in the $X$ channel. The curves are obtained for different $\theta$ from $\Phi_0$ to $\Phi_P$; (b) Normalized signal amplitude in the $X$ and $Y$ channels as a function of $B/G_0$ for $G_H/G_0=0.02$ and $\theta=\Phi_M$; (c) Normalized signal amplitude in the $X$ and $Y$ channels as a function of $G_H/G_0$ for $B/G_0=0.01$, and $\theta=\Phi_M$.

One important feature of the HMF regime is that one can distinguish quenching ($G_H<0$) from enhancing ($G_H>0$) signals, see Figure 4(c). As expected, the signal in the $X$ channel changes sign when the EDMR signal changes from enhancing to quenching, due to Eq. 9 and the even nature of the cosine function in Eq. 11. However, the signal in the $Y$ channel in the same situation does not change sign due to the odd nature of the sine function in Eq. 12.

Considering the above delineated features it could be possible to tune the modulation frequency, or the total circuit capacitance in order to allow the observation of the above-described HMF regime signal in the $Y$ channel and thus determine if the EDMR signal is quenching or enhancing. However, this procedure/interpretation must be carefully tested, since as discussed in Lee's work the phase response of the RSLs in EDMR depends on several parameters[26]. In addition it has also been observed by pulsed techniques that quenching and enhancing signals can be observed in the same sample on different time scales[14].

Note that measurements in the HMF regime, as described in Figure 4(a), could mask the signal due two RSLs with different phases, undermining the discrimination of the individual RSLs. This will become clear in the following section.

## IV. APLICATION TO EXPERIMENTAL DATA

In this section we apply our model to representative results of EDMR found in the literature. For this purpose, the fitting program *RESONA* was used, see Appendix B.

### A. Alq$_3$ based OLEDs

Figure 5(a) and Figure 5(b) show the EDMR spectra as observed in the $X$ and $Y$ channels of a PSD system for an undoped OLED at room temperature under the same experimental conditions but different modulation frequencies, in (a) $f$=13 Hz and (b) $f$=13 kHz. This device has the following structure Al/LiF/Alq$_3$/α-NPD/CuPc/ITO, details can be found in Ref. [25]. Figure 5(c) shows the quadrature EDMR spectrum from a DCM-TPA doped OLED at T=110 K with $f$=133 Hz. In this case, the device has the following structure Al/LiF/Alq$_3$/DCM-TPA/α-NPD/CuPc/ITO, see Refs. [25] and [33]. The red lines were obtained by fitting. The blue dashed line in Figure 5(b) was obtained using the fit parameters of Figure 5(a). Table 1 shows the parameters obtained experimentally and the best fits obtained using our model.

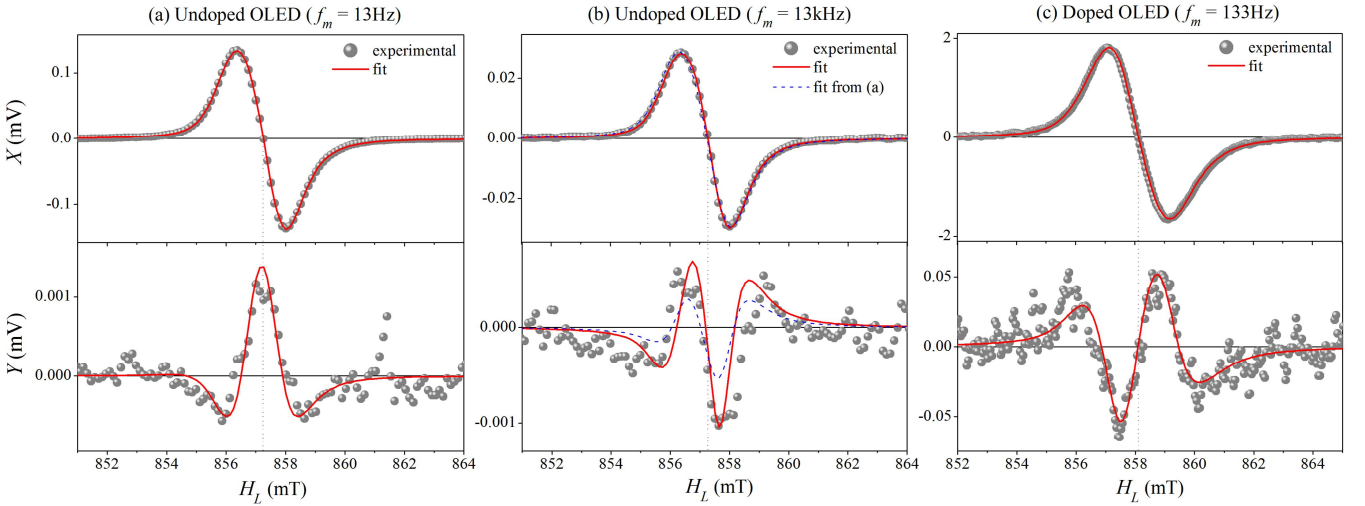

**Figure 5.** Simulation (red lines) of experimental EDMR spectra (gray points) for an undoped Alq$_3$ OLEDs using different magnetic field modulation frequency: (a) $f$=13Hz and (b) $f$=13kHz. Measurements were done at room temperature. In (c) the results using a doped OLED at 110K with $f$=133Hz are presented. The blue dashed line in (b) uses the same fitting parameters of (a).

**Table 1.** Fitting parameters from the present model (red line of Figure 5) compared to the values found in the literature. $\Delta H_{pp}$ were calculated from $\Delta H_{1/2}$*. $1/G_0$ values are obtained by dividing the applied DC voltage $V_0$ by the measured DC current $I_0$ corrected by the measuring resistance $R_M$. The errors were evaluated by averaging the result of 3 or more independent fitting procedures.

| Parameter | Undoped Alq$_3$ OLED | | | Doped Alq$_3$ OLED | |
|---|---|---|---|---|---|
| | Present work (13 Hz) | Present work (13 kHz) | Ref. 25 (133 Hz) | Present work (133 Hz) | Ref. 33 (133 Hz) |
| $g^a$ | 2.0028±0.0001 | 2.0030±0.0002 | 2.0028±0.0002 | 2.0026±0.0003 | 2.0039 |
| $g^b$ | 2.0045±0.0001 | 2.0044±0.0003 | 2.0040±0.0001 | 2.0049±0.0004 | 2.0042 |
| $\Delta g$ | 0.0017±0.0002 | 0.0013±0.0004 | 0.0012±0.0003 | 0.0023±0.0005 | 0.0003 |
| $\Delta H_{pp}^a$ (mT) | 1.03±0.05 | 1.00±0.02 | 2.0-3.4 | 1.51±0.05 | 1.82±0.01 |
| $\Delta H_{pp}^b$ (mT) | 1.10±0.03 | 1.11±0.03 | 1.5 | 1.46±0.05 | 3.0-4.2 |
| $\alpha^a$ | 0.14±0.11 | 0.20±0.06 | - | 0.16±0.07 | - |
| $\alpha^b$ | 0.61±0.05 | 0.48±0.10 | - | 0.51±0.12 | - |
| $G_H^a/G_0$ | -0.079±0.006 | -0.18±0.01 | - | 0.145±0.013 | - |
| $G_H^b/G_0$ | -0.055±0.003 | -0.11±0.01 | - | 0.154±0.010 | - |
| $\Delta\phi$ (°) | 1.2±0.2 | 1.3±0.5 | 0.4-2.8 [34] | 0.4±0.3 | - |
| $1/G_0$ (kΩ) | 45.7 | 42.6 | 45.7 | 790 | 790 |
| $R_M$ (kΩ) | 0.1 | 0.1 | 0.1 | 10 | 10 |
| $H_m$ (mT) | 0.5 | 0.5 | 0.5 | 0.5 | 0.5 |
| $C$ (nF) | 0 | 0.82 | - | 1.8 | - |

* $\Delta H_{pp} = \Delta H_{1/2}/\delta$, $\delta$ varying linearly from $\sqrt{3}$ (Lorentzian) to $\sqrt{2\ln 2}$ (Gaussian) considering $\alpha$ from Eq. 1.

Notice that the amplitude of the signal in the $X$ channel decreases strongly from $f$=13 Hz (Figure 5(a)) to $f$=13 kHz (Figure 5(b)), but there is no observable changes in the lineshape. The opposite occurs for the signal in the $Y$ channel: the maximum amplitude remains practically the same (~1 μV) but the lineshape changes significantly. Both features can be explained based on what was discussed in the previous section, for $f$=13 Hz (Figure 5(a)), $B/G_0$<0.01, thus the signal is in the LMF regime. From our model the signal in the $Y$ channel can only be explained by the contribution of two RSLs with different phases with respect to the reference. On the other hand, for $f$=13 kHz (Figure 5(b)), $B/G_0$~0.05, the measurements are in the HMF regime.

The strength of the proposed model can be better realized by its ability to describe EDMR spectra taken from the same device in two different modulation frequencies, 13 Hz in Figure 5(a), and 13 kHz in Figure 5(b), using the same fitting parameters. In fact the blue dashed line in Figure 5b is the EDMR signal predicted by the model from the fitted parameters on the experimental signal of Figure 5(a). It was obtained by using the overall capacitance of $C$=0.52 nF, which renders $\beta$=1 (no attenuation) for $f$=13 Hz and $\beta$=0.23 for $f$=13 kHz in Eq. 10. To get the best fit on Figure 5(b) (red curve), however, the overall capacitance was increased to $C$=0.82 nF, rendering $\beta$=0.11 for $f$=13 kHz, consequently the overall signal amplitude measured by the absolute value $G_H^a/G_0$ and $G_H^b/G_0$ has also to be increased for this case, see Table 1. This is an indication that the absolute values of $G_H/G_0$ can differ from the real variation of the sample's conductivity ($\Delta\sigma/\sigma$), as well as the capacitance value used to fit the data can differ from the real sample-plus-circuit capacitance, due to the simplicity of the equivalent circuit proposed. Moreover, it is well known that OLEDs cannot be completely described by simple RC circuits. In fact, impedance analysis show that $R$ and $C$ parameters present a significant dependence with frequency, including the existence of "negative capacitance" in low frequency regime[35-37].

As can be seen, the proposed circuit model is able to determine fundamental spectroscopic parameters of each RSL with relatively high precision and reliability. In Table 1 for the undoped OLED the values obtained from the fitting, except for $G_H/G_0$ values, are very close for the two frequencies under analysis. The new set of $g$-values is close to the earlier determined values of Ref. [25] for $f$=133 Hz. On the other hand the values for $\Delta H_{pp}$ obtained from the fits are systematically smaller than the experimental values in all cases. This could be related to the fact that $\Delta H_{pp}$ in the case of our model were calculated from the fitted $\Delta H_{1/2}$, while the data from the literature were determined directly from the experimental curves in the $Y$ channel using arbitrary phases. Moreover, the slight line broadening effect due the modulated field $H_m$, $H_m \sim \Delta H_{pp}/2$, is considered by the model. In fact, by ignoring this correction, the model returns $\Delta H_{pp}$ values 10 to 20% higher than those shown in Table 1. As discussed before, assuming that in Figure 5(b) the signal is in the HMF regime, the EDMR signal is found to be quenching, so the values for $G_H^a$ and $G_H^b$ are negative in Table 1.

For the doped OLED (Figure 5(c)), the $g$-values obtained from the fit are quite far from the previous reported values[33]. The $g$-difference ($\Delta g$) between the two RSLs is larger for the doped OLED ($\Delta g$=0.0023) than for the undoped OLED ($\Delta g$~0.0015). On the other hand, the phase difference ($\Delta\phi$) is higher for the undoped OLED $\Delta\phi$~1°, against $\Delta\phi$~0.2°.

However, it is worth to emphasize that, differently from $g$ and $\Delta H_{pp}$, $\Delta\phi$ is a dynamic parameter[26] and is probably strongly dependent on the experimental conditions. Thus, for a better discussion regarding $\Delta\phi$, we would need a systematic study varying experimental parameters such as temperature or microwave power. However, the relatively low values of $\Delta\phi$ reinforce the interpretation made in Refs. [22,25,33] that the EDMR signal in Alq$_3$ based OLEDs are due to one SDP with two spin pairs (electron and holes) experiencing slight different chemical environments. As mentioned earlier the transition from LMF regime to HMF is strongly dependent on the experimental conditions. In the case of the doped OLED, due to the low temperature used, a relative high resistance ($1/G_0$=790 kΩ) combined with a higher capacitance (1.8 nF) results in a HMF regime even at $f$=133 Hz. In this case, the comparison between the lineshapes of the signal at $X$ and $Y$ channels suggest an enhancing EDMR response ($G_H$>0) as revealed in Figure 4(c). As already pointed out in section III.B, in HMF the discrimination of the individual RSLs is undermined. This leads to imprecise determinations of $g$-factors and $\Delta H_{pp}$ values.

It is important to emphasize the relevance of the analysis considering both the regimes, LMF and HMF, which respectively allows the determination of spectroscopic parameters of distinct RSLs in the system and the nature of the dominating signal (quenching or enhancing). For example, the signals obtained in the HMF regime for undoped and doped OLEDS suggest the existence of distinct mechanisms in these devices responsible respectively for a quenching and enhancing EDMR signals. As shown by Lee *et al.*[26] it is very difficult to access the exact nature of the SPD responsible for each signal, however, based on a classical interpretation the quenching signal of undoped OLED is compatible to an increase on the singlet exciton population induced by ESR condition, while the enhancing signal of doped OLED could be associated to a detrapping processes[15].

### B. a-Si:H

In order to explore further the use of the circuit model, EDMR results from a-Si:H obtained by Dersch *et al.*[20] were analyzed. As already discussed, two independent SDPs were identified in a-Si:H[20]. The experimental quadrature EDMR spectra used in this section was built from published data, and thus the accuracy is limited compared to the previous section.

Figure 6 shows the reconstructed EDMR spectra in gray, taken from figure 6 of Ref. [20]. The solid red line shows the resulting fitting curves composed by the sum of the individual RSL shown as dashed blue and dotted green lines. Table 2 shows the parameters obtained from the fitting process.

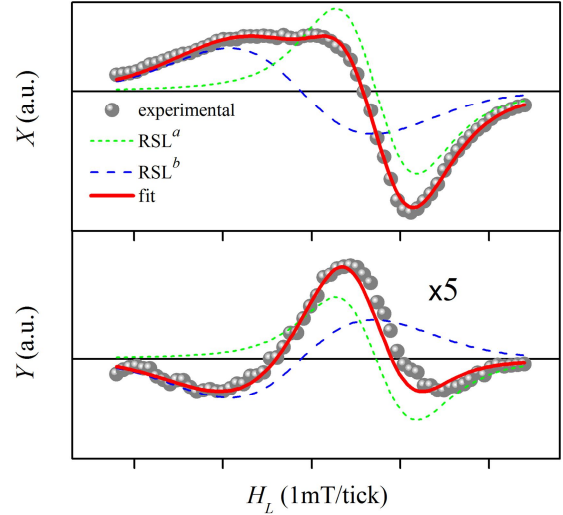

**Figure 6:** EDMR spectra of a-Si:H adapted from Ref. 21 shown in gray. The solid red line is the best fit using our model, together with its components, RSL$^a$ in dotted green and RSL$^b$ in dashed blue.

**Table 2.** Parameters used to fit the EDMR spectra from a-Si:H[20] using the proposed model as well as the corresponding published data.

| Parameter | a-Si:H | |
|---|---|---|
| | Present work | Ref. [20] |
| $g^a$ | 2.0051±0.0003 | 2.005 |
| $g^b$ | 2.0100±0.0005 | 2.01 |
| $\Delta H_{pp}{}^a$ (mT) | 0.69±0.05 | 0.6 |
| $\Delta H_{pp}{}^b$ (mT) | 1.51±0.09 | 1.2 |
| $\alpha^a$ | 0.15±0.05 | - |
| $\alpha^b$ | 0.75±0.11 | - |
| $\Delta\phi$ (°) | 16.6±0.9 | ~15 |

In this example, the fitting process was done with the limited experimental details available. For a more precise analysis, the complete set of experimental parameters is needed. Notice also that the model uses the lineshape of a mixture of symmetric Lorentzian and Gaussian functions, see Eq. 1, which do not correspond always to real situations, especially in inorganic materials.

Despite the limitations above described, Figure 6 clearly shows that the circuit model described here was able to fit the a-Si:H EDMR signal with a high degree of accuracy and the obtained values of the fitting parameters are quite close to those previous published, see Table 2.

The discussion of the results in Table 2 is not in the scope of this work, but it is interesting to notice the existence of a higher value of $\Delta\phi$, in comparison to those determined for Alq$_3$ based OLEDs. This discrepancy is compatible with the interpretation that a-Si:H signals are related to two SPDs, while Alq$_3$ OLEDs signals are associated just to one SPD with distinguishable precursor pair partners. Indeed weaker phase correlation could be expected in the response of RSLs coming from independent processes, allowing the observation of significant phases difference in the case of a-Si:H. On the other hand, given the stronger exchange interaction between electron-hole pairs smaller values of $\Delta\phi$ are expected for Alq$_3$ based OLEDs.

The origin of different phases between the SDPs or spin pairs is still not well understood. In our model it is a free parameter essential to describe the EDMR out of phase signal, however at this moment we are not able to describe its nature. In principle it could be associated to distinct relaxation times or another dynamic parameter related to transport/recombination processes, but more studies are still necessary in order to address more details about its physical origin.

In summary in this work we provide a circuit model that allows the extraction and evaluation of EDMR parameters with high accuracy and reproducibility. Two distinct regimes were identified depending on the magnetic field modulation frequency. These regimes define two distinct response domains, LMF and HMF, which carry complementary information about the system under study. In the LMF regime, the non-vanishing signal in the $Y$ channel is dominated by the dynamics of RSLs, allowing the isolation of relevant spectroscopic parameters, associated to distinct SDPs and/or spin pair partners. In the HMF regime, the signal in the $Y$ channel is dominated by the circuit capacitance, in such way that it is possible to discriminate situations where the EDMR signal is due to an increase or a decrease in the sample conductivity. In this context we point out the relevance of EDMR phase analysis and the regime in which the experiment is being performed, LMF or HMF.

Moreover, the proposed circuit model is constituted by simple mathematical equations, which allows its practical use in a fitting routine. In fact, a computational program called *RESONA* was developed in order to analyze and fit experimental quadrature EDMR spectra. The *RESONA* program, see Appendix B, includes unique features as line broadening correction due to over modulation, see Appendix A, and phase analysis by evaluating the entire quadrature EDMR signal in individual $X$ and $Y$ channels from PSD measurements.

## V. CONCLUSION

A model for EDMR based on an equivalent electrical circuit is presented and implemented. The proposed model is able to describe and predict a variety of complex quadrature EDMR signals with non-vanishing components in the $Y$ channel. Two different regimes have been described depending on the modulation frequency. In the low modulation frequency regime the model is able to isolate, discriminate and characterize individual contributions of distinct resonant spin lines. In the high modulation frequency regime, it is possible to determine whether the EDMR signal is quenching or enhancing.

The model was successfully applied to experimental data taken from the literature. The fitting process allows not only the determination of accurate spectroscopic values, as $g$-factor and $\Delta H_{pp}$ of the individual resonant lines, but also resolves the phase difference between them.

The simple equations used in the model allowed the development of a fitting program called *RESONA*, which can be easily implemented and are computationally fast. Typically it takes a few minutes to fit the data in a standard PC.

**Acknowledgements**

This work was financially supported by the Brazilian agencies CAPES (proc. 23038.008351/2010-17), FAPESP (proc. 2012/03116-7, 2011/21830-6 and 2008/57872-1 INCTMN) and CNPq. We are also thankful to Prof. F. Nüesch and Prof. L. Zuppiroli for providing the Alq$_3$ based OLEDs.

## APPENDIX A. LINE BROADENING DUE TO MAGNETIC FIELD MODULATION

The aim of this appendix is to provide a mathematical formula to simulate the line broadening effects due to magnetic field modulation on EDMR spectra. The approximations used are made in order to generate formulas that are easily implemented computationally.

From a mathematical point of view, in a lock-in amplifier the output signal is obtained by multiplying the input signal $V_M$ by the reference signal ($\sin\omega t$) and then integrating over the time constant *TC*. Considering this, $c_1$ in Eq. 9 can be written as:

$$c_1(H_L) = \frac{1}{TC} \int_{t_i}^{t_i+TC} G(t)\sin\omega t \, dt, \quad i = 0, 1, 2, \ldots, \Delta t/TC \quad (A1)$$

where the measuring time $t_i$ is related to $H_L$ by the following relation derived from Eq. 2:

$$H_L = H_0 + \Delta H\left(\frac{t_i + TC/2}{\Delta t} - \frac{1}{2}\right) \quad (A2)$$

For low speed field sweeps, $\Delta H/\Delta t \ll 2H_m f$, the variation of $H_L$ within *TC* can be disregarded, i.e., $H_L$ can be considered constant from $t_i$ to $t_i+TC$. Thus, $c_1$ can be approximated to:

$$c_1(H_L) = \frac{1}{TC} \int_{t_i}^{t_i+TC} G(H_L + H_m \sin\omega t)\sin\omega t \, dt \quad (A3)$$

where the phase difference between $G$ and the field modulation was set to zero, since we are only interested in the modulus of the signal.

Eq. A3 thus returns the effective or average slope of $G(H_L)$ within $H_L-H_m$ and $H_L+H_m$, which is supposed to be independent of *TC* when $TC \ll \Delta t$ and $TC = 2\pi k/\omega$ with $k=1,2,3,\ldots$ Therefore, choosing $k=1$, $TC=2\pi/\omega$, i.e., *TC* covers one single modulation cycle. Using $\xi=\omega t$, one can write:

$$c_1(H_L) = \frac{1}{2\pi} \int_0^{2\pi} G(H_L + H_m \sin\xi) \sin\xi \, d\xi =$$

$$= \frac{1}{2\pi}\left(\int_0^{\pi} G(H_L + H_m \sin\xi)\sin\xi \, d\xi + \int_{\pi}^{2\pi} G(H_L + H_m \sin\xi)\sin\xi \, d\xi\right) =$$

$$= \frac{1}{2\pi}\left(\int_0^{\pi} G(H_L + H_m \sin\xi)\sin\xi \, d\xi - \int_0^{\pi} G(H_L - H_m \sin\xi)\sin\xi \, d\xi\right) =$$

$$= \frac{1}{\pi}\int_0^{\pi/2}[G(H_L + H_m \sin\xi) - G(H_L - H_m \sin\xi)]\sin\xi \, d\xi$$

(A4)

In Eq. A4 $c_1$ is a continuous function of $H_L$ that can be numerically calculated by different commercial softwares. Another possibility is to use the following:

$$c_1(H_L) = \frac{1}{2n+1}\sum_{i=1}^{n}\left[G\!\left(H_L + H_m \sin\tfrac{i\pi}{2n+1}\right) - G\!\left(H_L - H_m \sin\tfrac{i\pi}{2n+1}\right)\right]\sin\tfrac{i\pi}{2n+1}$$

(A5)

where $n$ is chosen as the next integer of $2H_m/\Delta H_{1/2}$, i.e.: if $H_m/\Delta H_{1/2} \sim 1$, one uses $n=2$; if $H_m/\Delta H_{1/2} \sim 2$, uses $n=4$; and so forth.

In Figure A1 a comparison between various simulations are presented for different magnetic field modulation conditions. For $H_m/\Delta H_{1/2} \leq 0.2$, the EDMR signal amplitude is directly proportional to $H_m$, and there is no significant line broadening due to $H_m$. Thus for $H_m/\Delta H_{1/2} \leq 0.2$, Eq. 5 is a good approximation. However, for $H_m/\Delta H_{1/2} \geq 0.5$ shown in Figures A1(b-d), the signal amplitude is independent of $H_m$ and there is significant line-shape broadening. Thus, signal derived from Eq. 5 is no longer valid and the use of Eq. A4 or Eq. A5 are needed.

and (d) $H_m/\Delta H_{1/2}=2$. The open circles represent the signal using Eq. A1. The blue dashed line was obtained using Eq. 5, while the blue solid line Eq. A4. The green dash-dot line used Eq. A5 with $n=1$, while the red dashed line $n=4$.

It is clear from Figure A1 that the use of Eq. A5 with $n=4$ is a good approximation for $H_m \leq 2\Delta H_{1/2}$, and $n=1$ for $H_m < \Delta H_{1/2}$. The later has the following expression:

$$c_1(H_L) \approx \frac{\sqrt{3}}{6}\left[G\!\left(H_L + \frac{\sqrt{3}}{2}H_m \sin\tfrac{\pi}{3}\right) - G\!\left(H_L - \frac{\sqrt{3}}{2}H_m\right)\right]$$

(A6)

## APPENDIX B. THE FITTING PROGRAM – *RESONA*

The proposed circuit model described above is mathematically simple and allowed its implementation in a software named *RESONA*. The computational routine has a user-friendly interface allowing the operator to continuously modify each parameter and visualize the simulated and experimental data in two screens that represent the quadrature signal, $X$ and $Y$ channels, in real time, with the respective the sum of squared residuals.

*RESONA* is developed in Agilent VEE and is available for free at https://sites.google.com/site/resonaproject. To run the program it is only necessary to download and install the Agilent VEE Runtime program available for free at http://www.agilent.com.

The program has a single interface panel shown in Figure B1 that allows the input of the experimental parameters. Since commonly the magnetic field must be corrected, the program has this feature. All the simulation parameters are available using slider bars, and the main results are shown both graphically and in tables. A quick users guide is available by pressing the button "Info".

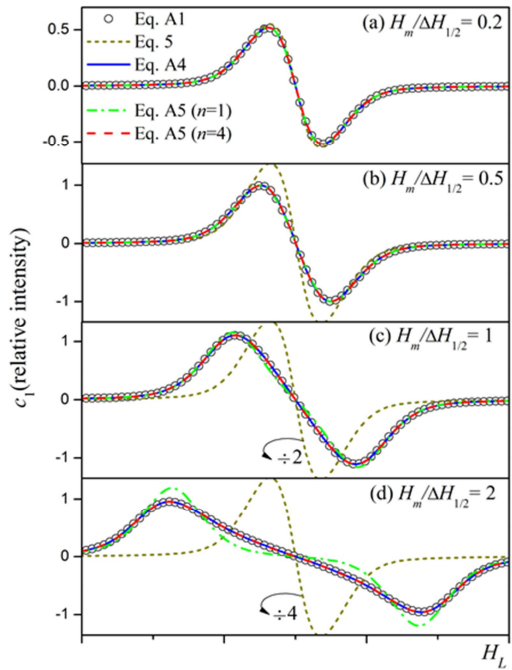

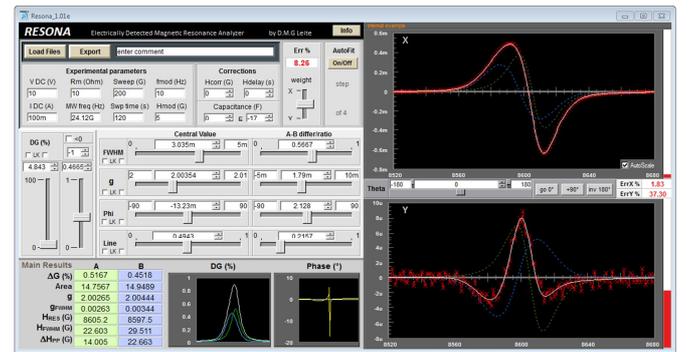

**Figure B1.** *RESONA* program interface with illustrative results.

The program has also an automatic fitting option that minimizes the sum of squared residuals of one or both graphs to get a fine adjustment of the modeling parameters. It takes 1 to 5 minutes to reach a good fit reproducibly. The values of the parameters employed and the resulting fitting curves can be saved. This program can also be used for ESR spectra simulation.

**Figure A1.** Simulated EDMR signal obtained from different equations of $c_1$ under different magnetic field modulation conditions: (a) $H_m/\Delta H_{1/2}=0.2$; (b) $H_m/\Delta H_{1/2}=0.5$; (c) $H_m/\Delta H_{1/2}=1$;